\documentclass[lettersize,journal]{IEEEtran}
\usepackage{amsmath,amsfonts}
\usepackage{algorithmic}
\usepackage{algorithm}
\usepackage{array}
\usepackage[caption=false,font=normalsize,labelfont=sf,textfont=sf]{subfig}
\usepackage{textcomp}
\usepackage{stfloats}
\usepackage{url}
\usepackage{verbatim}
\usepackage{graphicx}
\usepackage{cite}
\usepackage{soul,xcolor}
\usepackage{balance}

\begin{document}

\title{Advanced Architectures Integrated with Agentic AI for Next-Generation Wireless Networks}

\author{Kapal Dev,~\IEEEmembership{Senior Member,~IEEE,} Sunder Ali Khowaja,~\IEEEmembership{Senior Member,~IEEE}, Engin Zeydan,~\IEEEmembership{Senior Member,~IEEE}, Keshav Singh, Merouane Debbah,~\IEEEmembership{Fellow,~IEEE}
\thanks{Kapal Dev is with Munster Technological University, Cork, Ireland. Email: kapal.dev@ieee.org}
\thanks{Sunder Ali Khowaja is with School of Computing, Dublin City University and ADAPT Centre, Dublin, Ireland. Email: sunderali.khowaja@dcu.ie }
\thanks{Engin Zeydan is with Centre Tecnològic de Telecomunicacions de Catalunya, Castelldefels, Spain, 08860. Email: engin.zeydan@cttc.cat}
\thanks{Keshav Singh is with School of Computing, National Sun Yat-sen University, Taiwan. Email: keshav.singh@mail.nysu.edu.tw}
\thanks{ Merouane Debbah is with Khalifa University, Abu Dhabi, UAE. Email: merouane.debbah@ku.ac.ae}}



\maketitle

\begin{abstract}
This paper investigates a range of cutting-edge technologies and architectural innovations aimed at simplifying network operations, reducing operational expenditure (OpEx), and enabling the deployment of new service models. The focus is on (i) Proposing novel, more efficient 6G architectures, with both Control and User planes enabling the seamless expansion of services, while addressing long-term 6G network evolution. (ii) Exploring advanced techniques for constrained artificial intelligence (AI) operations, particularly the design of AI agents for real-time learning, optimizing energy consumption, and the allocation of computational resources. (iii) Identifying technologies and architectures that support the orchestration of back-end services using serverless computing models across multiple domains, particularly for vertical industries. (iv) Introducing optically-based, ultra-high-speed, low-latency network architectures, with fast optical switching and real-time control, replacing conventional electronic switching to reduce power consumption by an order of magnitude.

\end{abstract}


\section{Introduction}

The rapid development of telecommunications systems continues to drive the need for novel network architectures that meet the requirements of future technologies, especially in the transition to 6G \cite{raddo2021transition}. Unlike previous generations, 6G is set to revolutionize not only the performance and capacity of networks, but also enable new, intelligent services that support a variety of use cases, from immersive media to autonomous systems \cite{raddo2021transition}. 6G is also a communication and computing integrated platform as AI is considered to be its fundamental key enabler. The AI-integrated 6G system will be able to achieve higher level of automation, such as intent-based networking and automatic orchestration and maintenance (OAM). With the current evolution of AI towards large language models (LLMs) and Agentic approaches, the OAM can just simply issue intent for the network configuration to the AI agent, which then run the execution-feed-back-adjust-monitoring process along with the network tasks in an automated manner. Although, the Agentic AI is at its peak, the increasing complexity of the methods along with its support towards networks that are more scalable, efficient, and sustainable has not been discussed before. In response to these new challenges, the development of advanced network architectures integrated with Agentic AI should focus on several key aspects, such as, Reducing operational expenditure (OpEx), simplifying network operations, enabling dynamic service orchestration and integrating intelligent and autonomous functions. In this paper, we take a closer look at these architectural innovations and present solutions that go beyond the current limitations of 5G and early 6G frameworks.

One of the key innovations of this paper lies in the design of simplified network architectures that separate the control and user planes, allowing operators to deploy and manage services more efficiently through AI Agents while seamlessly integrating new network domains. This approach also addresses the limitations of existing service-based architectures (SBA) and provides a clear path for the long-term evolution of 6G networks. The increased flexibility of such architectures ensures that devices can operate as dynamic network nodes, enabling uninterrupted roaming across multiple operators and network technologies. In addition to architectural simplification, the integration of AI Agents into network management \cite{blanco2023ai} opens up new opportunities for optimization, especially in environments with energy constraints. Agentic AI-based networks can make intelligent decisions based on real-time data, but these models must work within strict energy and security constraints to ensure sustainable and secure operations. The concept of constrained AI operations, which is explored in this paper, is crucial for achieving these goals. It focuses on optimizing energy consumption and learning in real-time while maintaining security protocols. In addition, the emergence of serverless computing and function-as-a-service (FaaS) offers new solutions for the orchestration of services in distributed network environments. Serverless computing decouples application execution from the underlying infrastructure and enables real-time execution of lightweight functions at the network edge. This paradigm shift supports the flexible placement of AI agents and orchestration of tasks, a key requirement for the future 6G ecosystem where services must dynamically adapt to changing network conditions.

Another fundamental component of future 6G architectures is the use of autonomous cognitive agents that introduce decentralized decision-making processes \cite{wang2024survey}. These agents work without a central controller and enable dynamic and spontaneous interactions that optimize network resources and operations. This is a significant departure from traditional static architectures and enables more adaptable and resilient networks that can meet future requirements. The concept of goal-oriented communication, a revolutionary approach that enables AI-equipped devices to transmit only the information necessary to achieve specific goals \cite{strinati20216g}. This reduces unnecessary communication, computing load and energy consumption, while allowing devices to communicate more efficiently. Such innovations are crucial for next-generation machine-to-machine (M2M) communication, where optimizing resource usage is key. In addition, the potential for cloudification and management of RAN functions in space underscores the increasing role of distributed data centers, including those in space, in the dynamic orchestration of RAN functions \cite{liu2022cloud}. This approach contrasts with existing static methods of resource allocation and provides a more flexible and adaptable solution that can meet the diverse requirements of 6G communication scenarios. Cloudification of RAN functions is expected to play a critical role in achieving ultra- reliable, low-latency communications for a wide range of applications. The concept of Neural Radio Protocol Stacks (NRPS) can be used  to automate the generation of radio protocol stacks tailored to specific network requirements using  Agentic AI techniques. NRPSs can be a significant advance in the development of customizable and efficient communication protocols. By integrating Agentic AI into the development of radio stacks \cite{radiostack}, these solutions can offer a high degree of flexibility and adaptability, addressing issues such as interpretability, cross-vendor compatibility and computational efficiency.  The main contributions of the paper are as follows:
\begin{itemize}
    \item \textcolor{red}{We propose a novel 6G architectural framework that decouples the user and control planes and leverages Agentic AI for seamless cross-domain integration, enabling scalable and adaptive service deployment beyond the limitations of current SBA.} 
    \item \textcolor{red}{We introduce constrained Agentic AI techniques tailored for network operations, focusing on energy-efficient and secure real-time learning, with applications in dynamic resource allocation and threat-aware autonomous management.} 
    \item \textcolor{red}{We explore the integration of real-time serverless computing with Agentic AI for agile function orchestration, autonomous cognitive agents for decentralized network control, space-cloudified RAN, and neural radio protocol stacks for adaptive and efficient next-generation systems.}
    \item \textcolor{red}{We validate the effectiveness of Agentic AI through a V2X-based experimental study, demonstrating improvements in delivery packet ratio, latency, and link robustness compared to conventional methods.} 
\end{itemize}
    
The rest of the paper is organized as follows: Section \ref{agentic} discusses new 6G architecture models that leverage Agentic AI for flexible service provisioning, efficient resource management and real-time learning under energy and security constraints. It highlights serverless computing frameworks, decentralized cognitive agents for adaptive network management, goal-oriented communication protocols for bandwidth and energy efficiency, cloudification of radio access network (RAN) functions using space-based data centers, and customizable radio protocol stacks for real-time adaptation in 6G networks.  Section \ref{experimental} provides an experimental setup and an evaluation of agent-based AI within the vehicle-to-everything (V2X) scenario. Finally, Section \ref{conclusion} summarizes the key innovations discussed and outlines the key steps to realizing the potential of 6G networks, focusing on scalability, intelligence and efficiency.

\section{Agentic AIs}
\label{agentic}

\subsection{Novel Architectural Solutions for Simplified Networks with Agentic AI}

\begin{figure}[htp!]
\includegraphics[width=0.95\linewidth]{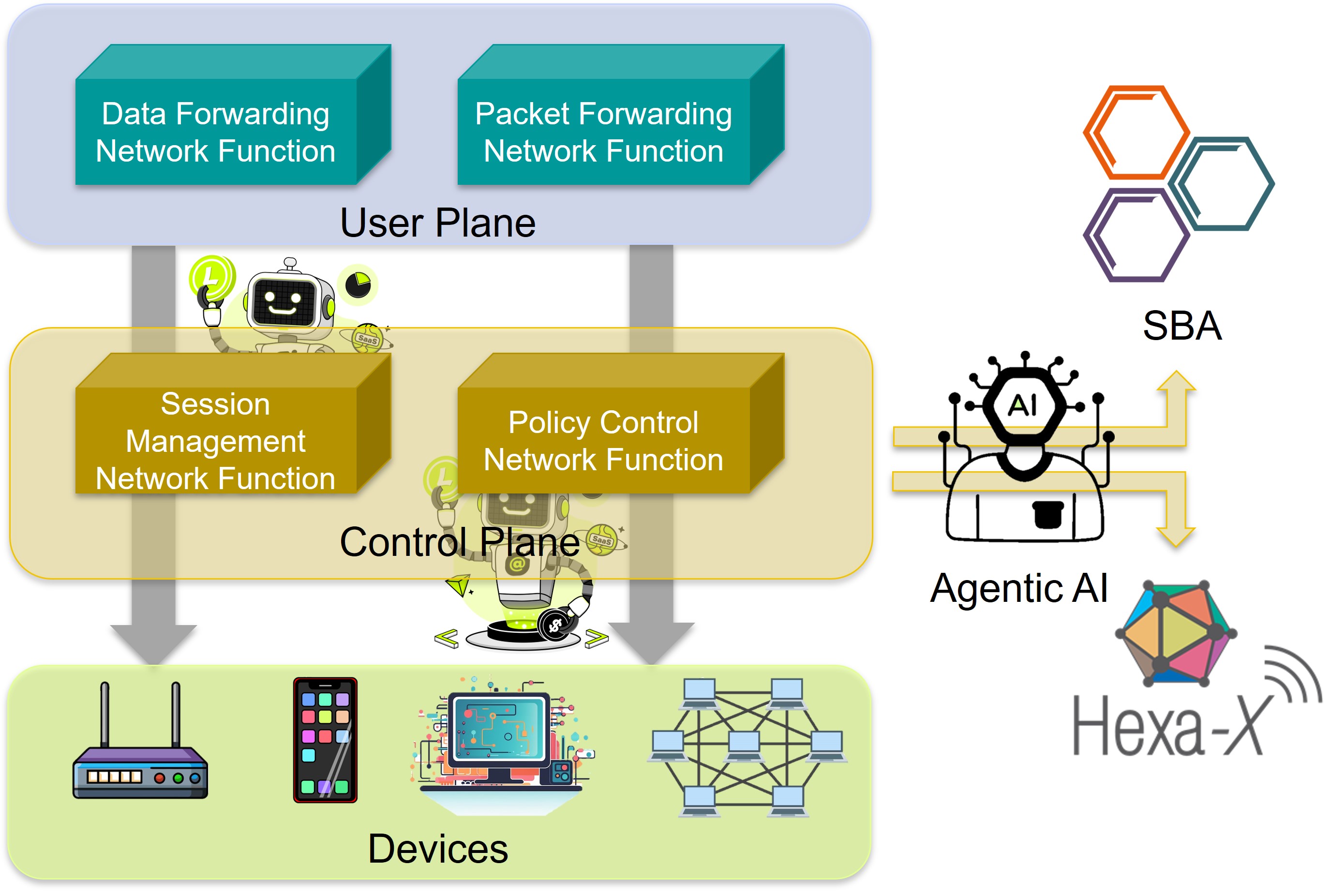}
\centering
\caption{A network diagram showing a simplified user-control plane architecture with seamless domain integration with Agentic AI}
\label{fig:new_arch}
\end{figure}

A primary objective in the design of future 6G networks is the simplification of network architectures to facilitate more efficient service deployment and operation using AI agents. 
\textcolor{red}{The limitations of the existing SBA in 5G networks are evident when considering the evolution toward 6G. Early models, such as the Hexa-X II framework \cite{hexa_x_ii_2024}, offer a foundation for addressing these shortcomings. However, further architectural refinements are needed to fully realize the potential of 6G, particularly with regard to supporting dynamic and adaptive services. The separation of the user and control planes is a foundational step toward flexible 6G architectures. AI agents play a critical role in dynamically managing this separation by continuously monitoring traffic patterns, resource demands, and network states. For instance, control plane agents can autonomously adjust routing policies, session management rules, and QoS parameters based on real-time analytics, while user plane agents optimize packet forwarding paths and load balancing. This decoupling allows for independent scaling and updating of each plane. Moreover, AI agents enable intent-based orchestration across both planes, like translating high-level service requirements into low-level configuration actions, thus reducing manual intervention while enhancing operational agility. Through semantic understanding and context-aware decision making, these agents ensure that control and user plane functions remain synchronized while adapting to heterogeneous network conditions. Fig. \ref{fig:new_arch} shows the network diagram  that uses simplified user-control plane architecture with seamless domain integration using Agentic AI, where AI agents will be designed to optimize the services, communication, flow of data using context, semantics, and sustainability characteristics. Hence, realizing the true potential of 6G.}

\textcolor{red}{From a standardization perspective, the separation of control and user planes with Agentic AI introduces new requirements for interface definitions and protocol specifications. Current 3GPP standards for SBA would need extensions to support AI-agent interactions, particularly for intent-based networking interfaces and cross-domain coordination. Ongoing initiatives in ETSI ISG F5G\footnote{https://www.etsi.org/technologies/fifth-generation-fixed-network-f5g} and ITU-T FG-NET-2030\footnote{https://www.itu.int/en/ITU-T/focusgroups/net2030/Pages/default.aspx} provide preliminary frameworks, but specific standards for AI-agent interoperability remain underdeveloped.} 



\subsection{Agentic AI for Constrained Network Operations}

\begin{figure}[htp!]
\includegraphics[width=0.95\linewidth]{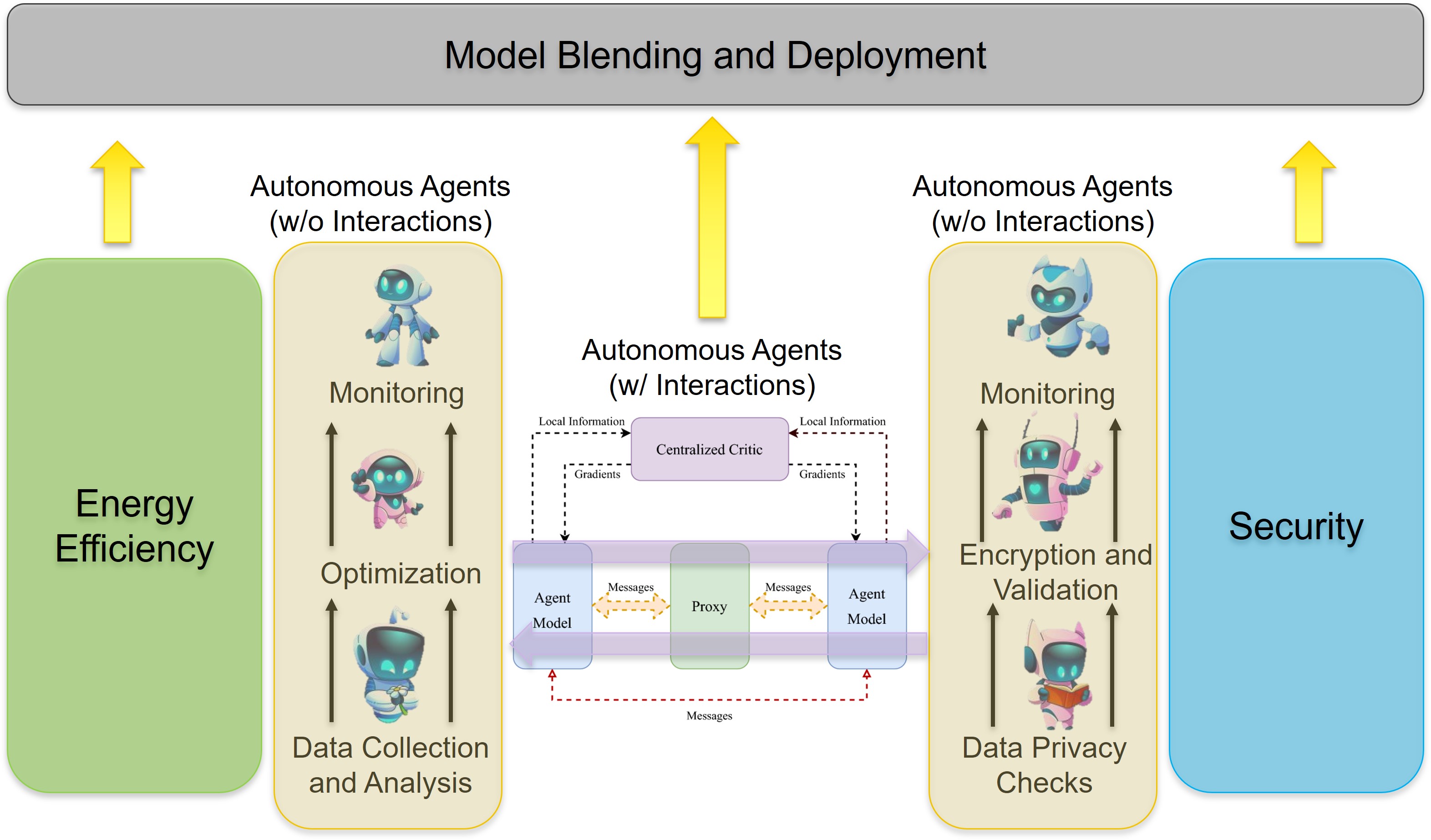}
\centering
\caption{A flowchart illustrating the process of deploying Agentic AI under energy and security constraints in network operations.}
\label{fig:Flowchart}
\end{figure}

Fig. \ref{fig:Flowchart} shows the flowchart for the Agentic AI implementation, showing a network optimized for both energy and security. The integration of Agentic AI into network operations offers significant optimization opportunities, especially in resource-constrained environments while adapting to new threats or energy constraints. Constrained AI refers to the application of predefined constraints, such as energy efficiency and security requirements, in the design and deployment of AI agents in networks. For example, AI agents can be assigned the task of collecting and analyzing data, optimizing energy efficiency and monitoring it sequentially. Same can be applied for the security constraints. However, these agents can also work together to make the decisions with semantics to integrate security and energy efficiency together. By incorporating these constraints into the learning process, it is possible to improve the performance of AI-driven network operations while ensuring compliance with domain-specific requirements.

In particular, the optimization of energy usage is a critical point when using AI agents in networks. Such a limitation also has an impact on the scalability of network operations, such as the support of massive machine type communication (mMTC). Energy-efficient AI solutions are essential for sustainable network operations, especially in the context of 6G, where the density of network infrastructure is expected to increase dramatically. In addition, the 6G networks must enable communication between space-terrestrial-ground- air, so the AI agents must be both energy efficient and computationally friendly. Furthermore, ensuring the security and robustness of AI agents in real-world conditions is of paramount importance as AI is increasingly used for autonomous network management and service provisioning. With the emergence of security vulnerabilities in AI models, concerns about the deployment of AI agents in 6G networks have grown \cite{SPIN}. Therefore, AI models must be resilient to specific AI model attacks such as model inversion, model poisoning, and membership inference attacks. As for communication between AI agents, security can be performed together with energy efficiency. For example, the autonomous agents can collect and analyze data, but instead of sending it for optimization, the security agent can perform privacy checks and encrypt the data. This will not only protect the data from intruders, but also learn to optimize the encrypted data. Also, the interactions between agents will help to understand the energy constraints so that privacy checks and encryption are performed using lightweight methods.

Although there is no specific information about reducing energy consumption in constrained AI operations, we can gain some relevant insights about AI-driven energy reduction in other contexts. Regarding AI-driven energy reduction in data centers, Google and DeepMind reported significant energy savings in data center cooling. DeepMind's machine learning system reduced the amount of energy used for cooling by up to 40\%. This resulted in a 15\% reduction in overall Power Usage Effectiveness (PUE) overhead after accounting for electrical losses and other non- cooling inefficiencies\footnote{https://deepmind.google/discover/blog/
deepmind-ai-reduces-google-data-centre-cooling-bill-by-40/}. Researchers at the MIT Lincoln Laboratory Supercomputing Center (LLSC) have developed techniques to reduce energy consumption when training AI models.
By implementing power capping on GPUs, they achieved a reduction in energy consumption of 12\% to 15\%. By using an early stopping technique, they were able to reduce the energy consumption for model training by a dramatic 80\% \cite{mit_ai_energy_2023}. With regard to AI-supported energy reduction in Radio Access Networks (RAN), Ericsson reported on energy savings in the operation of mobile networks. An ML/AI-based approach to dynamically configure cell sleep modes resulted in a 10-12\% energy reduction at pilot sites \footnote{https://www.ericsson.com/en/blog/2023/1/ai-powered-ran-energy-efficiency}.

\subsection{Real time serverless computing with Agentic AI for 6G Networks}
\label{serverless}

\begin{figure}[htp!]
\includegraphics[width=0.95\linewidth]{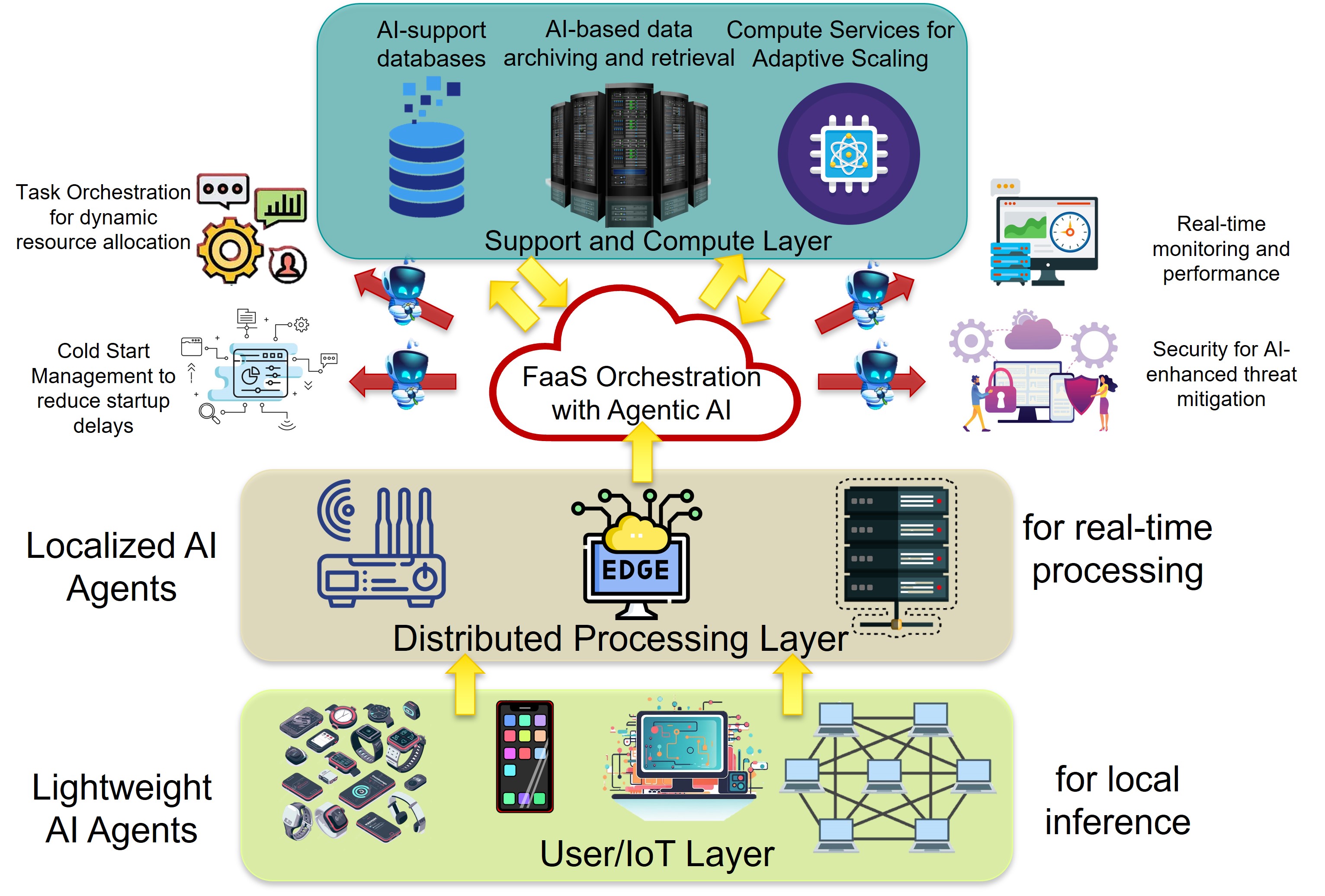}
\centering
\caption{A distributed network diagram showing the real-time execution of serverless functions with Agentic AI across edge and cloud environments.}
\label{fig:FaaS}
\end{figure}

Serverless computing, particularly in the form of FaaS  \cite{liu2023faaslight}, offers a promising solution for the dynamic orchestration of network services while employing Agentic AI as shown in Fig. \ref{fig:FaaS}. Achieving seamless scaling is difficult in existing network architectures due to the accessibility issues. By decoupling the execution of services from the underlying infrastructure, serverless computing enables the deployment of lightweight, highly-scalable functions across the network capable of automating and optimizing the task, including at the edge. In the context of 6G integrated with Agentic AI, real-time serverless computing can address several key challenges. These include mitigating the cold start problem, orchestrating high-dimensional tasks across multiple network domains, and ensuring security in distributed computing environments. The serverless computing with AI agents can also address the seamless scaling issues by employing proactive resource allocation and function pooling. The ability to dynamically allocate resources based on real-time network conditions is critical to achieving the low-latency, high-reliability performance required in next-generation networks. The latency of serverless computing FaaS with Agentic AI in real time can vary greatly depending on the platform, complexity of the function and network conditions. As detailed in \cite{moreno2023latency}, cold start latency can range from 100 milliseconds to several seconds. In a typical machine learning application for image classification, cold start latencies of about 2-3 seconds. Warm start latency, which occurs when an agent is called for a particular function shortly after a previous call, is typically in the range of 10-100 milliseconds  and in a cloud scenario where a device sends image files to an AWS Lambda function deployed in a remote Amazon region, network latencies were between 70 and 120 milliseconds. In an edge scenario, where the device sends files from the same region as the Lambda function, network latencies were reduced to 0.5 to 2 milliseconds.

\subsection{Autonomous Cognitive Agents in Network Architectures}
\label{autonomous}

The use of autonomous cognitive agents in network architectures represents a significant shift from the traditional, centralized network management models. Multi-Agent Systems (MAS), potentially using large language models (LLM), enable decentralized decision making where individual agents interact and cooperate to achieve network goals \cite{zeeshan2024large}. While it can be argued that the LLMs require computationally intensive operations, which could hinder the realization of the said network architecture, various techniques such as low-rank adapters (LoRA), quantization and others can be used to mitigate the computational complexity issues of the LLMs while reducing the computational complexity of the system. These agents can dynamically invoke each other, enabling a flexible and adaptable network infrastructure that operates in a more organic and decentralized manner. The design of architectures that support autonomous agents requires careful consideration of scalability, security and the seamless integration of service composition and knowledge handling. Decentralizing network operations not only improves efficiency, but also increases the network's ability to adapt to changing conditions, providing robustness and resilience.

\begin{table*}[htp!]
\centering
\scriptsize
\caption{Comparison of 6G Enablers: Characteristics, Pros, and Cons}
\begin{tabular}{|p{2.5cm}|p{4cm}|p{4cm}|p{4cm}|}
\hline
\textbf{6G Enabler} & \textbf{Characteristics} & \textbf{Pros} & \textbf{Cons} \\ \hline
\textbf{Novel 6G Architectures} & 
Separation of User and Control planes using AI agents, Flexible network management and resource allocation, and autonomous network nodes & 
- Simplifies network operations \newline
- Seamless integration with new network domains \newline
- Uninterrupted user roaming & 
- Additional interfaces \newline
- Management Overhead \newline
- More read and store stateful data \\ \hline

\textbf{Agentic AI for Constrained Network Operations} & 
Energy-aware AI, Security-driven learning, Real-time optimization algorithms for network processes & 
- Reduce energy consumption 
\newline
- Sustainable network operations \newline
- Secure and Resilient system & 
- High computational requirements \newline
- Latency trade-offs in complex models \newline
- Sensitive to model privacy attacks \\ \hline

\textbf{Real-time Serverless Computing with Agentic AI} & 
Dynamic function orchestration, Edge execution, Function placement at different network layers (cloud to edge) & 
- Reduces latency to as low as 10 ms with warm start \cite{moreno2023latency} \newline
- Seamless scaling \newline
- Support for high-dimensional tasks across multiple network domains & 
- Cost of long-running processes \newline
- Lack of custom control \newline
- Difficulty in Function Management \\ \hline

\textbf{Autonomous Cognitive Agents} & 
Decentralized decision-making, Multi-agent systems (MAS), Knowledge sharing, LLM-driven coordination & 
- Hypercustomization \newline
- Improved fault tolerance (20\% increase) \newline
- Reduces dependency on centralized control & 
- Stable Training and Tokenization \newline
- Security vulnerabilities (Membership Inference Attacks) \newline
- Seamless Integration \\ \hline
\textbf{Neural Radio Protocol Stacks with Agentic AI} & 
ML-based automated protocol generation, Customizable and efficient protocol stacks, Cross-vendor compatibility & 
- Optimized protocol stacks for specific use cases \newline
- Performance Enhancement \newline
- Enables real-time protocol adaptation & 
- Challenges in interpretability and explainability \newline
- Hardware constraints at the radio edge \newline
- Limited real-world PoC demonstrations \\ \hline
\end{tabular}
\label{tab:comparison}
\end{table*}

\subsection{Neural Radio Protocols Stacks (NRPS) with Agentic AI}

\begin{figure}[htp!]
\includegraphics[width=0.85\linewidth]{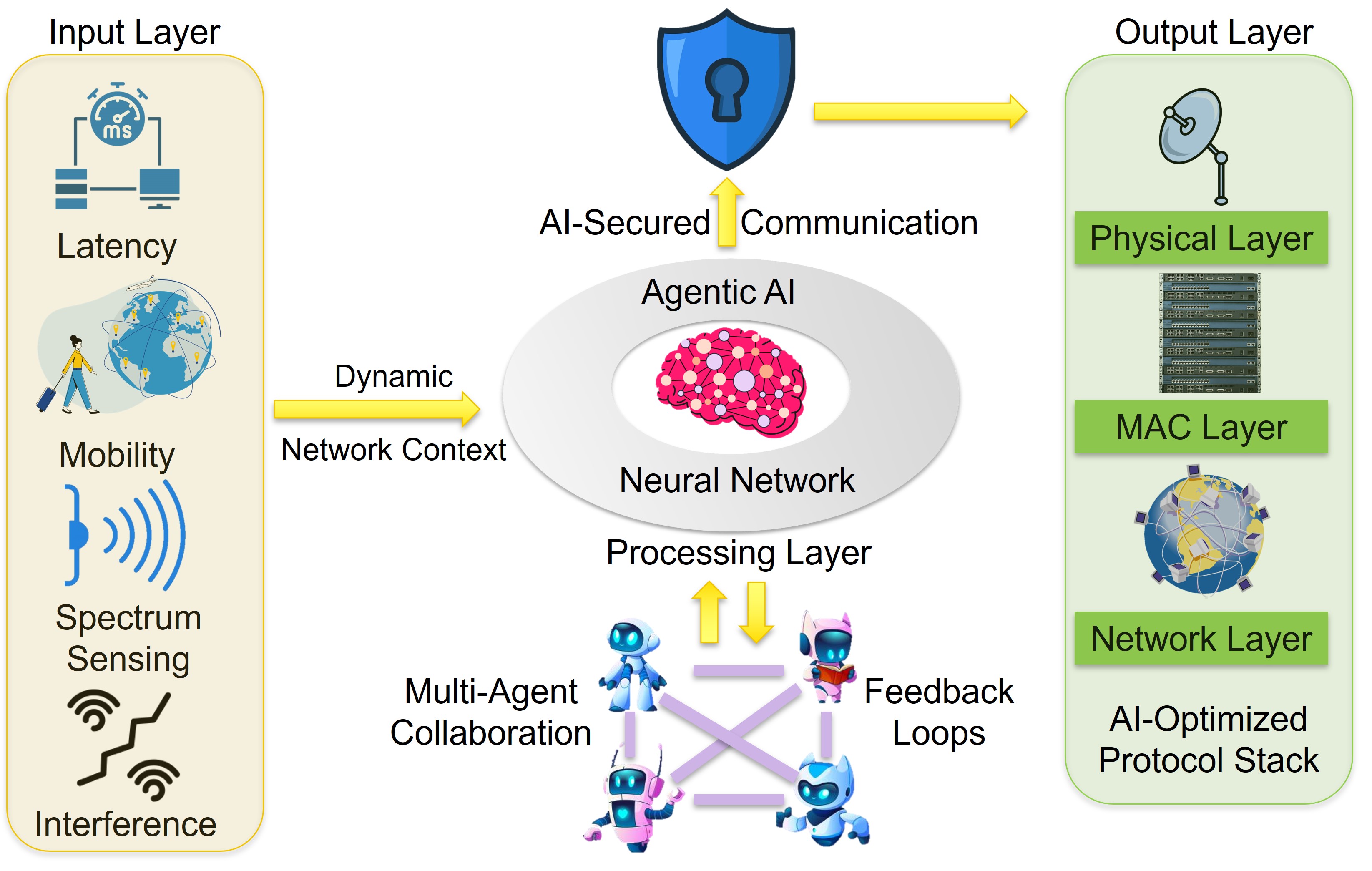}
\centering
\caption{A diagram showing the automatic generation of neural radio protocol stacks using neural Agentic AI, focusing on interpretability and efficiency.}
\label{fig:neural}
\end{figure}

The concept of Neural Radio Protocol Stacks (NRPS) with Agentic AI uses autonomous AI agents to adapt, manage and optimize radio communication protocols in real time. The NRPS help to adapt to dynamic network conditions while collaboratively optimizing the communication networks. The use of Agentic AI for NRPS will also enable autonomous decision-making for spectrum sensing and allocation, dynamic protocol adaptation, cooperative communication and multi-agent collaboration, security and anomaly detection, energy-efficient communication, and learning and evolving protocols \cite{racz2022full}.
This introduces machine learning-based techniques to the design and customization of radio protocol stacks as shown in Fig. \ref{fig:neural}. By using neural networks, it is possible to automatically create lean, adaptive protocol stacks that are optimized for specific network conditions. However, there are still some challenges, including ensuring the interpretability of neural stacks, achieving cross-vendor compatibility, and optimizing computational efficiency.

The development of NRPS with Agentic AI requires not only advances in algorithmic techniques, but also innovations in the hardware architectures that can support the deployment of such stacks at the radio edge. Proof-of-concept (PoC) demonstrations are essential to validate the feasibility and performance of Agentic AI-enabled NRPS in real-world scenarios. Recently, however, AI chips have been announced that promise to provide Agentic AI-based services on the edge devices. Similarly, model compression techniques and edge-optimized software frameworks could help in implementing the concept of Agentic AI-based NRPS to achieve the desired adaptation of radio protocol stacks. 

\textcolor{red}{Lets consider the domain of channel state information (CSI) compression in order to discuss the practical applicability of NRPS. Traditional CSI feedback mechanisms are often inefficient under dynamic channel conditions. With NRPS, an AI agent can learn to compress CSI feedback adaptively by selecting only the most informative features based on real-time channel characteristics, thereby reducing overhead while maintaining link reliability. Similarly, in spectrum sharing scenario, NRPS-enabled agents can autonomously negotiate and allocate spectral resources among multiple operators by modeling spectrum usage patterns and interference constraints. For instance, in a shared license/unlicensed band scenario, agents can dynamically adjust transmission parameters such as power and modulation to maximize coexistence efficiency, leveraging real-time sensing and multi-agent collaboration.}
Finally, Table \ref{tab:comparison} provides a comparison of the 6G enablers using Agentic AI as discussed in this paper along with their features, advantages and disadvantages.

\section{Experimental Setup and Evaluation}
\label{experimental}

To validate Agentic AI for next-generation network architectures, we conduct a simple experiment to optimize packet delivery ratio and connection robustness in a V2X environment. \textcolor{red}{The emulation of V2X environment in our experimental setup leveraged 100 to 500 devices moving at speeds between 10-30 m/s under moderate to high interference conditions. Node placement was randomized and clustered to simulate urban and highway scenarios. We implemented three specialized AI agents using LLaMA2-13B and LLaMA2-70B models, each optimized for performance, sustainability, and spectral efficiency. These agents were integrated via a crewaI framework\footnote{https://github.com/crewAIInc/crewAI} and supported by a Retrieval Augmented Generation (RAG) system built from 60 ArXiv papers, Telecom Q\&A datasets \cite{TelecomQA}, and 3GPP documentation. The V2X environment for vehicle mobility was simulated using simulation of urban mobility (SUMO)\footnote{https://sumo.dlr.de/docs/Downloads.php} and network emulation was simulated using NS-3. We used the scripts from the repository\footnote{https://github.com/addola/NS3-HelperScripts/tree/master/examples/SUMOTraceExample} for SUMOTrace, to gather the results. We used the socket connection to communicate between the Python agent code and the NS-3 simulation.} The agents collaboratively adjusted transmission parameters such as power control, modulation schemes, and retransmission threshold in real time. When multiple AI agents collaborate, they can tackle complex problems that would overwhelm a single agent working alone. By sharing what they know and dividing up tasks, the agents create a multiplier effect, i.e. their combined intelligence lets them solve challenges more effectively than any one agent could on its own. Think of it like a well-coordinated team where each member brings their unique strengths to the table. 

\textcolor{red}{The experimental setup was designed to emulate 5G/6G usecase, i.e. URLLC for vehicular safety communications in V2X environment. The aforementioned setup assumes that the scenario models the transmission of cooperative awareness messages (CAMs) and decentralized environmental notification messages (DENMs) between vehicles and infrastructure, which are essential for collision avoidance and cooperative driving. The parameters are also configured according to 3GPP Release 16 and ETSI specifications for NR-V2X, including sub-6 GHz band operation, 10-100 MHz channel bandwidth, and slot-based transmission. The mobility patterns generated in SUMO reflect realistic urban scenarios with vehicle densities. The alignment of the experimental setup with standardized 5G/6G use case represents a real-world network deployment.} 

To make it fair and simple, we did not apply any pre-processing or specialized curation of information to train AI agents, other than feeding the RAG. We modified the method proposed in \cite{V2X} and performed an evaluation without the use of AI agents. 
We evaluate performance based on the packet delivery ratio (PDR), i.e. the percentage of packets successfully delivered, and the robustness of the connection, i.e. the mean signal-to-noise ratio (SNR) and retransmission rate. Note that we repeat the simulation for a different number of parameters and give average results accordingly. Table \ref{table2} shows the comparative analysis for the parameter optimization of PDR and Link Robustness in V2X with and without Agentic AI. \textcolor{red}{The results demonstrate that Agentic AI outperforms conventional methods across all metrics. The improvement in PDR (95\% vs. 79\%) and mean SNR (30dB vs. 20dB) highlights the agents' ability to adapt to dynamic channel conditions. The reduction in retransmission rate (5\% vs. 17\%) and latency (20ms vs. 50ms) underscores the efficiency of AI-driven parameter tuning. These gains are attributed to the agents' collaborative reasoning, which enables proactive resource allocation and failure prediction. The use of RAG further allowed the agents to leverage domain knowledge without extensive retraining, illustrating the practicality of lightweight AI integration into operational networks.} 

\begin{table}[htp!]
\centering
\caption{Comparative analysis for parameter optimization of PDR and Link Robustness in V2X with and without Agentic AI. }
\label{tab:my-table}
\begin{tabular}{|c|c|c|}
\hline
\textbf{Metric}              & \textbf{Without AI Agents} & \textbf{With AI Agents} \\ \hline
\textbf{PDR}                 & 79\% ($\pm$ 5\%)            & 95\% ($\pm$ 2\%)         \\ \hline
\textbf{Mean SNR}            & 20 dB ($\pm$ 2dB)          & 30 dB ($\pm$ 1dB)       \\ \hline
\textbf{Retransmission Rate} & 17\%                       & 5\%                     \\ \hline
\textbf{Adaptability}        & Low                       & High                  \\ \hline
\textbf{Latency}             & 50 ms ($\pm$ 10 ms)        & 20 ms ($\pm$ 5 ms)      \\ \hline
\end{tabular}
\label{table2}
\end{table}

\section{Open Issues, Challenges, and Future Directions}
\textcolor{red}{Although the integration of Agentic AI into 6G networks presents several benefits and promises to overcome issues related to existing iterations of communication system, it has still unresolved challenges that require immediate research attention. We highlight some of open issues concerning Agentic AI based 6G networks and provide associated future directions to address those issues. \\
\emph{Real-time Orchestration of Serverless Functions}: Current FaaS platforms struggle with the cold-start latencies exceeding 100ms, which is unacceptable for URLLC scenarios. This latency bottleneck is exacerbated by the stateless nature of serverless functions when maintaining AI agent context across multiple invocations. Future research should explore predictive function warming algorithms using temporal pattern analysis and stateful serverless containers that preserve agent memory across executions, potentially reducing cold-start times to under 10ms, especially for mission-critical applications.\\
\emph{Interpretability and Trustworthiness of NRPS}: While AI-generated protocols can dynamically adapt to channel conditions, their black-box nature raises concerns about operational transparency and compliance with regulatory requirements. The inability to explain why specific protocol parameters were selected hinders adoption in safety-critical systems like autonomous vehicle communications. Future works should investigate hybrid symbolic-AI approaches that combine neural networks with human-curated RAG systems, creating auditable decision trails while maintaining adaptive capabilities. Developing standardized testing frameworks for NRPS verification across different vendor implementations will also be essential for commercial deployment.\\
\emph{Sustainability and Energy Usage for Distributed AI agents}: The energy sustainability poses a fundamental constraint for widespread 6G deployment. As AI models grow in complexity, their computational demands conflict with the energy efficiency goals of green networking. Current LLMs require significant processing power, creating a carbon footprint that undermines network sustainability objectives. Research directions should focused on specialized TinyLLMs for edge devices, and renewable-energy-aware agent scheduling that aligns computational workloads with available green energy sources.\\
\emph{Multi-vendor Interoperability}: Standardization gaps represent a critical barrier to multi-vendor interoperability in Agentic AI networks. The absence of unified interfaces for agent-to-agent communication and agent-to-network function interaction creates proprietary silos that limit scalability. Current 3GPP specifications lack definitive standards for AI agent authentication, service level agreements, and conflict resolution mechanisms. Future standardization efforts should prioritize open API specifications for agent communication, reference architectures for AI-native networks, and certification frameworks that ensure cross-vendor compatibility.\\
\emph{Security vulnerabilities in Multi-agent Systems}: The security aspect require urgent attention to mitigate vulnerabilities in Agentic AI systems. The distributed nature of Agentic AI creates expanded attack surfaces, including model poisoning with malicious agents, adversarial attacks on decision processes, and privacy breaches through inference attacks. Traditional security mechanisms are insufficient for detecting sophisticated threats against collaborative AI systems. Future research should develop byzantine-resistant aggregation techniques for distributed learning, explainability-based anomaly detection that identifies aberrant agent behavior, and homomorphic encryption methods that allow secure computation on encrypted network data without compromising agent functionality. \\
\emph{Standardization of Agentic AI protocols}: Standardization represents a critical pathway for the adoption of Agentic AI in 6G networks. Key standardization bodies including 3GPP, ITU-R, and IETF will need to address several dimensions, which include interface standardization between AI agents and network functions, certification frameworks for AI model safety and security in critical communications, performance benchmarking methodologies for AI-driven network operations, and ethical guidelines for autonomous network decision-making. The timeline for these standardization efforts aligns with 3GPP Release 19-20 timeframe, where initial studies on AI/ML for networks are already underway. Collaboration between academic researchers, industry stakeholders, and standard organization needs to be established in order to develop a coherent framework and comprehensive standards for Agentic AI network integration that enables innovation while ensuring interoperability and security.}

\section{Conclusion}
\label{conclusion}

In this paper, we provide a comprehensive examination of the key architectural advances that will shape the future of 6G networks when integrated with Agentic AI. From novel 6G architectures to Agentic AI for constrained network operations, serverless computing with Agentic AI and neural protocol stacks with Agentic AI, the innovations discussed in this paper are expected to transform the telecommunications landscape, enabling networks that are smarter, more efficient and capable of supporting a wide range of new applications. Key contributions such as novel architectural solutions that simplify network operations through the decoupling of the user and control planes through autonomous AI agents, and introducing Agenting AI-based constrained optimization techniques to reudce energy consumption and improve security in real-time network operations. Real-time serverless computing with Agentic AI offers a flexible approach to dynamic function orchestration, while autonomous cognitive agents enable decentralized and adaptive network management. 

Additionally, we explored the Neural Radio Protocol Stacks using Agentic AI that highlights the potential for customizable and efficient radio protocols in future networks. We conducted an experiment for parameter optimizing in the context of V2X with and without AI agents. We designed 3 AI agents to provide optimized parameters recursively based on the current state and evaluated the performance using PDR, link robustness, retransmission rate, adaptability, and latency. We show that the parameter optimization with Agentic AI is far better than the conventional techniques. 

\textcolor{red}{As 6G standardization activities gain momentum in organizations such as 3GPP, ITU-R, and IEEE, the integration of Agentic AI principles into future specifications will be crucial. Our work provides technical foundations that can inform these standardization efforts, particularly in the areas of network architecture, constrained AI operations, and dynamic resource optimization. The experimental validation demonstrates the tangible benefits of Agentic AI approaches, supporting their consideration in upcoming 6G standards.} 

\balance

\bibliographystyle{ieeetr}
\bibliography{biblio}

\begin{thebibliography}{10}

\bibitem{raddo2021transition}
R.~Singh, A.~Kaushik, W.~Shin, M.~D. Renzo, V.~Sciancalepore, D.~Lee, H.~Sasaki, A.~Shojaeifard, and O.~A. Dobre, ``Towards 6g evolution: Three enhancements, three innovations, and three major challenges,'' {\em IEEE Network}, vol.~Early access, pp.~1--10, 2025.

\bibitem{blanco2023ai}
L.~Blanco, S.~Kukli{\'n}ski, E.~Zeydan, F.~Rezazadeh, A.~Chawla, L.~Zanzi, F.~Devoti, R.~Kolakowski, V.~Vlahodimitropoulou, I.~Chochliouros, {\em et~al.}, ``Ai-driven framework for scalable management of network slices,'' {\em IEEE Communications Magazine}, vol.~61, no.~11, pp.~216--222, 2023.

\bibitem{wang2024survey}
L.~Wang, C.~Ma, X.~Feng, Z.~Zhang, H.~Yang, J.~Zhang, Z.~Chen, J.~Tang, X.~Chen, Y.~Lin, {\em et~al.}, ``A survey on large language model based autonomous agents,'' {\em Frontiers of Computer Science}, vol.~18, no.~6, p.~186345, 2024.

\bibitem{strinati20216g}
T.~M. Getu, G.~Kaddoum, and M.~Bennis, ``Semantic communication: A survey on research landscape, challenges, and future directions,'' {\em Proceedings of the IEEE}, vol.~112, no.~11, pp.~1649--1685, 2024.

\bibitem{liu2022cloud}
B.~Agarwal, R.~Irmer, D.~Lister, and G.-M. Muntean, ``Open ran for 6g networks: Architecture, use cases and open issues,'' {\em IEEE Communications Surveys \& Tutorials}, vol.~Early Access, pp.~1--37, 2025.

\bibitem{radiostack}
J.~Wang, W.~Jiang, R.~Liu, and S.~Wang, ``Towards efficient and portable software modulator via neural networks for iot gateways,'' {\em IEEE Transactions on Mobile Computing}, vol.~23, no.~12, pp.~13866--13881, 2024.

\bibitem{hexa_x_ii_2024}
``{Hexa-X-II: The European 6G Flagship Project}.'' \url{https://hexa-x-ii.eu/}, 2024.
\newblock Accessed: 2024-10-13.

\bibitem{SPIN}
S.~A. Khowaja, P.~Khuwaja, K.~Dev, and A.~Antonopoulos, ``Spin: Simulated poisoning and inversion network for federated learning-based 6g vehicular networks,'' in {\em ICC 2023 - IEEE International Conference on Communications}, pp.~6205--6210, 2023.

\bibitem{mit_ai_energy_2023}
A.~Boutouchent, A.~Mekrache, A.~Ksentini, G.~Adhane, J.~P. C.~d. Fonseca, J.~McNamara, K.~Ramantas, M.~Palena, M.~Iordache, R.~L. Cigno, S.~Mostafa, S.~Roy, and C.~Verikoukis, ``6g-intense: Intent-driven native artificial intelligence architecture supporting network-compute abstraction and sensing at the deep edge,'' {\em IEEE Vehicular Technology Magazine}, vol.~20, no.~1, pp.~44--54, 2025.

\bibitem{liu2023faaslight}
X.~Liu, J.~Wen, Z.~Chen, D.~Li, J.~Chen, Y.~Liu, H.~Wang, and X.~Jin, ``Faaslight: General application-level cold-start latency optimization for function-as-a-service in serverless computing,'' {\em ACM Transactions on Software Engineering and Methodology}, vol.~32, no.~5, pp.~1--29, 2023.

\bibitem{moreno2023latency}
R.~Moreno-Vozmediano, E.~Huedo, R.~S. Montero, and I.~M. Llorente, ``Latency and resource consumption analysis for serverless edge analytics,'' {\em Journal of Cloud Computing}, vol.~12, no.~1, p.~108, 2023.

\bibitem{zeeshan2024large}
T.~Zeeshan, A.~Kumar, S.~Pirttikangas, and S.~Tarkoma, ``Large language model based multi-agent system augmented complex event processing pipeline for internet of multimedia things.'' arxiv preprint, https://arxiv.org/abs/2501.00906, 2025.

\bibitem{racz2022full}
F.~Jiang, C.~Pan, L.~Dong, K.~Wang, O.~A. Dobre, and M.~Debbah, ``From large ai models to agentic ai: A tutorial on future intelligent communications.'' arxiv preprint, https://arxiv.org/abs/2505.22311, 2025.

\bibitem{TelecomQA}
L.~Bariah, H.~Zou, Q.~Zhao, B.~Mouhouche, F.~Bader, and M.~Debbah, ``Understanding telecom language through large language models,'' in {\em IEEE Global Communications Conference (GLOBECOM)}, pp.~6542--6547, IEEE, 2023.

\bibitem{V2X}
V.~Todisco, S.~Bartoletti, C.~Campolo, A.~Molinaro, A.~O. Berthet, and A.~Bazzi, ``Performance analysis of sidelink 5g-v2x mode through an open-source simulator,'' {\em IEEE Access}, vol.~9, pp.~145648--145661, 2021.

\end{thebibliography}

\vfill

\end{document}